\documentclass[runningheads]{llncs}
\usepackage{graphicx}
\usepackage[
backend=biber,
style=numeric,
sorting=ynt
]{biblatex}
\begin{document}
\title{Using Network Analysis on Twitter Data to Identify Threats on Indonesian Digital Activism}
%
%
\author{Adya Danaditya}
%
%
\institute{School of Computer Science, Carnegie Mellon University, 
\\Pittsburgh, Pennsylvania 15213, USA\\
\email{adanadit@andrew.cmu.edu}}
\maketitle              
\begin{abstract}
In this study, we tried to see and characterize potential threats to digital activism in the internet-active nation of Indonesia by doing network analysis on a recent digital activism event on Twitter, which protested against a recent law related to alcoholic beverage investment. We hoped insights from the study can help the nation moving forward as public discourses are likely to stay online post-COVID. From this study, we found that threats in form of hashtag hijackings happen often in digital activism, and there were traces of a systematic information campaign in our observed case. We also found that the usage of bots is prevalent in and they showed significant activity, although the extent to which they influenced the conversation needs to be followed through more. These threats are something to think about as activism goes increasingly digital after COVID-19 as it can imbue unwanted messages, sow polarization, and distract the conversation from the real issue.

\end{abstract}
\section{Introduction}
In 1990, a first glimpse of digital activism happened when people protested through email and online forums against Lotus MarketPlace, which set out to sell personal information of potential customers in America to interested buyers \cite{mccaughey_2014}. From there, the online world has played a major role in activism – it was credited to be a major force in the 2011 Arab Spring \cite{alsayyad_guvenc_2013} and an organizing point in the Occupy Wall Street movement \cite{al-hasan_yim_lucas_2019}. While not every digital activism is effective and resulted in a positive real-world outcome \cite{christensen_2011},  people still actively participate in such acts. A study suggested 53\% of Americans engaged in digital activism and political engagements in social media in 2018 \cite{anderson_toor_rainie_smith_2018} and there are no significant signs that suggests that the trend will stop.

The recent physical restrictions imposed on the general populace due to the COVID-19 pandemic posed a challenge to traditional ways of activism in the issue of form and continuity, but at the same time, the crisis also primed traditional activism for a radical change and further transition to digital formats \cite{activism}. The year of the pandemic, interestingly, also coincided with tons of trigger events that sparked social movements like Black Lives Matter and Stop Asian Hate \cite{rosenblatt} - and with it we can see the rise of digital activism to mainstream consciousness. All these facts make digital activism a very current and important topic to explore at this day and time.

In this study, we try to characterize potential threats in the realm of digital activism during the COVID-19 pandemic in the country of Indonesia as a way to understand what lies ahead in this space. The country is a hotbed of Twitter activity - It boasts the 6th largest Twitter population in the world, with 14.9 million users tweeting per January 2021 \cite{tankovska_2021}. Activism is also deep in the country's psyche. The current form of democracy was basically set up by student protests (“Reformasi”) after the 32-year semi-authoritarian Orde Baru (New Order) and people were not shy on protesting policies they dislike openly: street movements were regularly held to protest against disparate policies from as a new set of investment and labor law being passed \cite{omnibus} to rise in gas prices \cite{bbm}.

In the following sections, we look at an recent activism event on Twitter and analyze what kinds of visible threats emerged on that case through network analysis. The activism event revolves around a policy by the president regarding investment on alcoholic beverages: The Indonesian president Joko Widodo had issued a decree to take out liquor and alcoholic beverage-related investments from the investment blacklist in several key Indonesian cities in order to boost foreign investment \cite{tempo}. This decision is met with public furor, especially from the conservative Muslim groups in the country, and he decided to scrap the decree shortly afterwards \cite{tempo_revoke}. We hope the findings and insights can guide the country to navigate online discourse in future Indonesian digital activism situations.
\section{Literature Review}
In our literature review, we look for common forms of threats known to digital discourse to guide our analysis in this study
\subsection{Hashtag Hijacking}
Hashtags are a tool used in most social media platforms to structure topics being discussed by users in their platform \cite{laucuka_2018}. Due to its key role in how users find topics to engage in social media, they are commonly used as a rallying point for digital advocacy and activism \cite{freelon_mcilwain_clark_2016}. These facts entice malicious users to exploit hashtags as a way to steer conversations elsewhere by piggybacking the growing trend, a practice commonly known as ‘hashtag hijacking’ \cite{vandam_tan_2016}. The practice of hashtag hijacking has been widely studied and is known to be prevalent in political discourse through social media \cite{hadgu_garimella_weber_2013}.

\subsection{Information Operations}
Social media has been frequently used as a front of conflict between state actors or non-state actors \cite{singer_brooking_2019}. It has been used also as tool for propaganda to change social behaviors or beliefs for political gains \cite{woolley_howard_2019}. We would co-opt the term ‘information operations’ to characterize actions that lie in between these two spheres. 
The emergence of these information operations poses a new threat for national security, and that necessitates the emergence of a new field of study in order to respond to it. 

Social cybersecurity is a branch of applied computational social science that seeks to characterize threatening cyber-mediated social changes and tries to build infrastructures to counter that exact threat \cite{carley_2020}. Contributions to this field give us a glimpse of the tell-tale signs of an information operation within a social network, and in this study, we would use the BEND framework, which groups specific actions to common objectives in information campaigns \cite{carley_2019} as a way to detect information operations within our observed activism events.

\subsection{Bot Usage}
Bots, which are automated or semi-automated accounts that are designed to mimic human behavior \cite{bot}, are commonly found in conjunction with information operations to amplify conflict or polarization and also to help bridge seemingly unconnected communities for nefarious purposes \cite{carley_2020}.
Studies have shown that bots are prevalent in political discourses all across the world. Specific studies indicated bot activity for varying objectives and degrees of success in social media discourse surrounding important elections in some Asian countries \cite{uyheng_carley_2019}\cite{uyheng_ng_carley_2021} and a publication by Oxford Internet Institute has specifically shown that bots are prevalent in political discourse in Indonesia - and even might be used actively by the government to support policies, attack opposition and drive divides in the society \cite{oii}.

\section{Methodology}

\subsection{Data Collection}
The study begins with collecting Twitter data for the observed case. We used hashtags related to the observed topic as the query and we got those hashtags by observing trending topic lists around the time the topic are known to arise in the public consciousness. With the hashtags (which include \#TolakInvestasiMiras, \#BatalkanPerpresMiras, \#PapuaTolakInvestasiMiras and \#MirasPangkalSejutaMaksiat), we proceeded in fetching the tweets related to them with Twitter's V2 Search API. 

\subsection{Data Processing}
There was significant processing done to the JSON response from Twitter's API to ensure compatibility with ORA. Other processing efforts include annotating the network with results from Bothunter - a supervised random forest model with multi-tiered feature additions that predicts a Twitter account's probability of being a bot. \cite{bothunter}. The annotation is done to help us understand bot activity in the observed network.

\subsection{Network Analysis}
ORA \cite{Carley2017} is used to derive networks from the Twitter data, and from we use reports and visualizations that it generated to perform analysis of the network. The reports and visualizations used in this study include: 
\begin{itemize}
\item Calculation of network measures over time after aggregating the Twitter data per day
\item Iterative visualization of the network graph of co-occurring hashtags and communication between accounts (mentions, quotes, and retweets)
\item Network grouping locator based on the Louvain algorithm \cite{traag_waltman_eck_2019} done on the co-occurring hashtag network to detect possible hashtag hijackings
\item Detection of key entities based on common network node measures like degree centrality and betweenness centrality in the communication network between accounts
\end{itemize}

\subsection{Qualitative Analysis}
The visualization and the report results from ORA are then matched with the insights we gained from the literature review. We summarized the content of the tweets, the behavior of the accounts and used it to guide how we should interpret the findings from the network in conjunction with the literature review.

\section{Findings}

\subsection{Hashtag Hijacking}
We identified multiple cases of hashtag hijacking in our case. These cases of hashtag hijacking are relatively vulgar in practice - most tweets just discussed totally different matters with different and jarringly co-opt the activism hashtag, as shown in this sample, which ties the activism hashtag to K-Pop related hashtags: 
\textit{"Not everything will go as you expect in your life. This is why you need to drop expectations, and go with the flow of life. \#MinMarch \#SUGA
\#YoongGi \#MinYoongi \#BTS \#ARMY \#TolakInvestasiMiras \#TolakLegalisasiMiras"}

We also find that by doing a Louvain grouping \cite{traag_waltman_eck_2019}, an algorithm commonly used to find clusters and groups over a network, on the co-occurring hashtag network we could roughly group hashtags by the kind of topic that they talked about - and this is a promising premise to detect hashtag hijacking. As seen in Table 1, there are samples of resulting groups from the Louvain grouping that clearly show other topics that accounts try to steer the discourse of the hashtag into.

\begin{table}
\centering
\caption{Louvain grouping result samples - alcoholic beverage investment policy activism.}\label{tab3}
\begin{tabular}{|l|l|} 
\hline
\multicolumn{2}{|l|}{\textbf{Groups Highlighting Hashtag Hijackings}}                                                                                                                                                                                      \\ 
\hline
\textbf{\textbf{Prevalent themes}}                                             & \begin{tabular}[c]{@{}l@{}}\textbf{\textbf{Topic samples in group }}\\\textbf{\textbf{or component hashtags}}\end{tabular}                                                \\ 
\hline
\begin{tabular}[c]{@{}l@{}}Indonesian opposition \\talking points\end{tabular} & \begin{tabular}[c]{@{}l@{}}Calls to impeach the president \\and his party, vice president role, \\Corruption of COVID relief fund, \\protest of Omnibus Law\end{tabular}  \\ 
\hline
Reference to K-Pop fandoms                                                     & \begin{tabular}[c]{@{}l@{}}BTS and Bayern 3 debacle, \\\#ARMY,~ \#Shinee, \#Suga, \\\#WeSupportYouJennie~ ~ ~ ~ ~~\end{tabular}                                           \\ 
\hline
Random 4 character hashtags                                                    & \begin{tabular}[c]{@{}l@{}}fa17, 5cn2, 6l35, ffn1\\U3o2, 350e, jc3h, etc\end{tabular}                                                                                     \\
\hline
\end{tabular}
\end{table}

With this method, we also found one group that is very peculiar - as it seems to group together hashtags that are exclusively random 4 characters, e.g. \#fa17, \#5cn2, \#6l35. We investigate this group deeper in the next section.

\subsection{Evidence of Information Operation}
The random character hashtag group we found in the previous section contains 3591 hashtags, and 94\% of those hashtags are only mentioned once in the network. We also found that the agents that tweeted those hashtags are surprisingly well connected to each other in the communication network. Isolates only account for 10\% of the whole agent nodeset and the network of these agents have a high measure of hierarchy and connectedness.

The tweets tweeted out of this group mostly consist of canned messaging with thousands of micro-variations accompanying the peculiar hashtags. This messaging contains calls that appeal to Islamic values to reject the bill and/or show sympathy to the notion of Khilafah, or an Islamic state in Indonesia (both independent and in conjunction with the previous point). Some of these tweets are posted with mentions to @YuanaRyanTresna, a cleric who has a considerable following online. The sample of canned messages used can be seen in Table 2. 

The actions of the accounts in this hashtag group, including the closely-knit communications and attempts to engage a possible influencer, can be frame them very neatly to the framework the BEND maneuver provides to characterize information operation - The ‘positive’ spectrum section of the BEND maneuver to manipulate social networks can help us structure and rationalize these found acts as follows:

\begin{itemize}
\item \textbf{Build} : This short hashtag tweeting actors are mentioning each other

\item \textbf{Back} : YuanaRyanTresna are engaged, building his influence – sample tweet: \\
\textit{@Su*** @Ro**** @YuanaRyanTresna @Yo** Meski keadaan ekonomi negeri ini defisit, tidak lantas hrs melegalkan sesuatu yang ALLOH SWT haramkan \#MirasPangkalSejutaMaksiat \#MirasIndukMaksiat \#zyl7 (Despite the economic deficit, we don't need to condone something that God forbid)}

\item \textbf{Bridge} : Connection is setup between YuanaRyanTresna’s audience and the bots – this is done by enticing YuanaRyanTresna to retweet one of the incoming mentions from the short hashtag tweeting accounts

\item \textbf{Boost} :
Khilafah philosophies are sent as a possible solution to the chaos created by the alcoholic beverage debacle – sample tweet: \\
\textit{@Za**** @Ros**** @YuanaRyanTresna @Yo**** Persoalan Negeri Tak Kunjung Selesai: Apakah Tegaknya Khilafah Solusinya? \#4c8f (There's no end to this country's problem, is Khilafah the solution?)}
\end{itemize}

\begin{table}
\caption{Canned messaging samples.}\label{tab2}
\centering
\begin{tabular}{|l|l|} 
\hline
\textbf{\textbf{Templates}}                                                                                                                                                                                                                                                                                                                                                                                                                                                                                                                                                      & \textbf{Sample Tweets}                                                                                                                                                                                                                                                                        \\ 
\hline
\begin{tabular}[c]{@{}l@{}}\textless{}Measure of time\textgreater{}, \\ \textless{}Alhamdulillah /MasyaAllah\textgreater{} \\saya melihat di kota Jakarta sudah \\banyak orang yang sadar khilafah, \\kamu gimana \textless{}mention twitter account\textgreater{}\\\\\textbf{\textbf{\textit{Rough translation:~}}} \\ \textless{}Alhamdulillah/MasyaAllah\textgreater{} \\I see that there’s a lot of people in Jakarta \\this last \textless{}Measure of time\textgreater{} \\ that is aware of Khilafah. \\How about you? \textless{}mention twitter account\textgreater{}\end{tabular} & \begin{tabular}[c]{@{}l@{}}2 tahun terakhir, Alhamdulillah saya melihat \\ di kota Jakarta sudah banyak orang \\ yang sadar Khilafah. \\ kamu gimana @ka****\\\\4 bulan terakhir, MasyaAllah saya melihat \\ di kota Jakarta sudah banyak orang \\ yang sadar Khilafah. \\ kamu gimana @Che****\end{tabular}  \\ 
\hline
\begin{tabular}[c]{@{}l@{}}\textbf{\textbf{Same message multiple forms:}}\\ \textit{Tying criminal cases with alcohol use} \\ -\textgreater{} Maraknya kasus kriminal tersebab miras\\-\textgreater{} Di daerah, kasus kriminalitas yang \\ disebabkan miras juga marak.\\\textbf{\textbf{\textit{}}}\\\textbf{\textit{Rough translation:~}} \\ -\textgreater{}The abundance of criminal cases\\ is because of alcohol\\\textbf{\textbf{-\textgreater{}~}}In many regions, alcohol-related \\ cases are also prevalent\end{tabular}                                                       & \begin{tabular}[c]{@{}l@{}}@r***** Maraknya kasus kriminal \\ tersebab miras\\@n***** Maraknya kasus kriminal \\ tersebab miras\\@s****~ Di daerah, kasus kriminalitas yang \\ disebabkan miras juga marak.\\@U****~ Di daerah, kasus kriminalitas yang \\ disebabkan miras juga marak.\end{tabular}      \\ 
\hline
\multicolumn{1}{l}{\begin{tabular}[c]{@{}l@{}}\\~ ~ ~ ~ ~ ~~\end{tabular}}                                                                                                                                                                                                                                                                                                                                                                                                                                                                                                       & \multicolumn{1}{l}{}                                                                                                                                                                                                                                                                         
\end{tabular}
\end{table}

These findings would suggest a strong possibility of an organized information operation sympathetic to the Khilafah movement existing in the alcoholic beverage investment case. This might be done by the operators in hopes of piggybacking the heavy religious sentiments appearing in the activism movement, in turn being able to spread their message to a broader Islamic audience.

\subsection{Bot Prevalence}
Bothunter is used to identify bots amidst the accounts in the network and their prevalence. We used a threshold score of .75 on the output to base the judgment on whether the account is bots or not. The result of this processing showed that bot activity is prevalent in our observed case.

We observed that 3586 accounts, or 21.12\% of all of the accounts in the network are bots, with the rest, again, being either not bots or cases in which accounts that are verified, has been suspended or has been suspended since. The accounts that are involved in the suspected information operation scored high in bot activity – on that group, bots account for 23.99\% of all accounts, and 37.78\% percent of the accounts are either suspended or deleted since, possibly due to it being identified as malicious or bot-like by Twitter, leaving only 38.23\% to be non-bots. This is aligned with the previously stated premise that bots are widely used in information operations. 

The impact that these bots caused to the activism, apart from the discursive impacts of the information operation that they try to support, is quite profound - if we take the bots from the random character hashtag group earlier in isolation, for example, they managed to take the hashtag ‘\#MirasPangkalSejutaMaksiat’ on to the Indonesian trending topic chart on February 28th due to their activity.

\section{Discussions}
\subsection{Impact and assessment of the practice of hashtag hijacking}
While we found lots of co-occurring hashtags seemingly trying to steer conversations to other topics, we found that as a result there are no cluster of tweets big enough talking about those topics to suggest any success in hashtag hijacking attempts. 

However, through identification of hashtag hijackings, we found the random-four character hashtag group related to a suspected information operation. This group, in conjunction with the aspects put forth by the suspected information operation itself, would be much more profound effect to actors participating in this activism, as elaborated further in the next subsection. 

\subsection{Information operations on Islam-related political topics}
The finding of a BEND-like maneuver in the activism event of the alcoholic beverage investment is significant, as it's a clear evidence of an attempt to expose the predominantly devout Muslim opponents of the alcoholic beverage investment policy to Islamic state propaganda. We can see how they try to connect the narrative between the excesses in society alcoholic beverages can bring and the hardline stance a possible Islamic state could do to solve that. This might affect regular citizens and possibly drag them to a more extreme position on the matter of an Islamic state as a result, though the exact impact and motivation of the group in doing this is less clear and need to be researched further.

\subsection{The role of bots}
We find that bots are a legitimate threat in both our cases by the fact that they are prevalent and they are influential in carrying out information operations. Prevalence means that the chance that an organic account would pass them as legitimate and picked up their narrative is quite high. Also, it is demonstratively shown in the case of the alcoholic beverage investment that their mobilization could be used to amplify trends at specific points in time. This could be a strategic advantage in the domain of propaganda and information warfare.

\subsection{Limitations and future directions}
There are several limitations of this study – firstly the dataset itself: while hashtags provide a neat grouping of tweets related to a certain topic, there is no way to ascertain for sure that the set is comprehensive, there might be other tweets related to the topic that is not covered. Also, while we can show that there are strong indicators of an information operation creeping in the event of activism observed it’s hard to authoritatively determine the underlying motivations of this operation with the data we collected. A further literature review or deeper research can help enhance analysis on this matter. Most importantly, we might need to further look to see whether the patterns and findings present in this study holds in other observations of digital activism events in Indonesia.

\section{Conclusions}
At this point we have identified possible threats to look out for in Indonesian digital activism. These include topic hijackings, information operations and bot infiltration. From our observations, we saw that hashtag hijackings attempts happen often, but the intensity and nature of it varies from topic-to-topic: we found a systematic information campaign in the more political case as well as other sporadic, more random and diverse set of hijacking attempts. Bots are prevalent, to varying degrees of impact, although the extent to which they influence the conversation needs to be followed through more. As a general note, this study suggests hijackings, bots and information operations are a clear threat Indonesians need to be aware of if activism goes increasingly digital after COVID-19 as it can be weaponized, imbue unwanted messages, sow polarization, and distract the conversation from the real issue.

\subsubsection{Acknowledgment} This paper is adapted from a class project for Dynamic Network Analysis course delivered by Professor Kathleen Carley. The writer would like to thank Dr. Carley, Jeremy Straughter, and Lynnette Ng, for all the inputs that all of them has contributed on the development of this paper. The author also would like to thank Lembaga Pengelola Dana Pendidikan, the institution that funded the author’s study at Carnegie Mellon University.

%
%
%
%

\printbibliography

\end{document}